\documentclass[pre,amsmath,amssymb,floatfix,twocolumn,showpacs]{revtex4}
\usepackage{graphicx}
\usepackage{float}
\usepackage{bm}
\usepackage[utf8]{inputenc}

\begin{document}
\title{Reply to Comment on Effect of polydispersity on the ordering transition of adsorbed self-assembled rigid rods}
\author{N. G. Almarza}
\affiliation{Instituto de Qu{\'\i}mica F{\'\i}sica Rocasolano, CSIC, Serrano 119, E-28006 Madrid, Spain }
\author{J. M. Tavares}
\affiliation{Centro de F\'{\i}sica Te\'orica e Computacional, Universidade de Lisboa, Avenida Professor Gama Pinto 2,
P-1649-003 Lisbon, Portugal}
\affiliation{Instituto Superior de Engenharia de Lisboa, Rua Conselheiro Em\'{\i}dio Navarro 1, 
P-1950-062 Lisbon, Portugal}
\author{M. M. Telo da Gama}
\affiliation{Centro de F\'{\i}sica Te\'orica e Computacional, Universidade de Lisboa, Avenida Professor Gama Pinto 2,
P-1649-003 Lisbon, Portugal}
\affiliation{Departamento de F\'{\i}sica, Faculdade de Ci\^encias, Universidade de Lisboa, Campo Grande,
P-1749-016 Lisbon, Portugal}

\begin{abstract}
We comment on the nature of the ordering transition of a model of equilibrium polydisperse rigid rods, on the square lattice, which is reported by L\'opez et al. to exhibit 
random percolation criticality in the canonical ensemble, in sharp contrast to (i) our results of Ising criticality for the same model in the grand canonical ensemble [Phys. Rev. E {\bf 82}, 061117 (2010)]
and (ii) the absence of exponent(s) renormalization for constrained systems with logarithmic specific heat anomalies predicted on very general grounds by Fisher [M.E. Fisher, Phys. Rev. {\bf 176}, 257 (1968)].
\end{abstract}

\date{\today}
\maketitle

Extensive Grand Canonical Monte Carlo simulations, for a model of adsorbed self-assembled rigid rods (SARR) on the square lattice, indicate that the polydisperse rods undergo a continuous transition in the two-dimensional (2D) Ising class, in line with models of monodisperse rods \cite{SARR,Matoz}. This finding is in sharp contrast to a previous result, 
based on Canonical Monte Carlo simulations, where equilibrium polydispersity was claimed to change the nature of criticality, from Ising to random percolation \cite{Lopez}.

In the preceding comment L\'opez et al. elaborate on this claim to conclude that the criticality of the SARR model on the square lattice depends both on polydispersity and 
on the statistical ensemble. This surprising result is based on simulations and normal finite-size scaling analysis of the SARR model in the canonical and grand canonical ensembles. 
The conclusion was that the SARR model exhibits random percolation criticality ($q=1$ Potts) when the system is in the canonical ensemble, while the criticality is Ising-like 
($q=2$ Potts) when the system is in the grand canonical ensemble. This is at odds with very general arguments by Fisher \cite{Fisher} on the absence of exponent renormalization in constrained (e.g. fixed density) systems with logarithmic specific heat anomalies, as well as with the results of a detailed simulation study of Fisher scaling of the 2D Ising 
magnetic lattice-gas, by Ferreira and Prodanescu \cite{ALF}. Fisher has also shown that although the universality class of constrained systems, with specific heat anomalies, is 
that of the unconstrained ones there are logarithmic corrections to the scaling functions, which may affect the scaling behaviour of reasonably sized systems as shown by Ferreira 
and Prodanescu \cite{ALF}.  

The existence of two universality classes for the SARR model, claimed by L\'opez et al., is based on the calculation of (i) the fourth-order Binder cumulant of the order parameter, 
$\delta$, $g_4=1-<\delta^4>/(3<\delta^2>^2)$ at the transition, $g_4^{c}$, and (ii) the value of the correlation length exponent $\nu$, obtained by normal scaling data collapse of 
the cumulants, for different system sizes. Both the values of $g_4^c$ and $\nu$, reported by L\'opez et al. for the SARR model, in the canonical and grand canonical ensembles, are different.

In the canonical simulations of the preceding comment L\'opez et al. kept the surface coverage constant and varied the temperature of the system, rather than fixing the temperature and varying the coverage \cite{Lopez}. As discussed 
below the logarithmic corrections to the normal finite-size scaling analysis, arising from the constant density constraint apply in both cases. In other words, Fisher logarithmic corrections \cite{Fisher,TSARR} as well as the simpler logarithmic correction suggested by us \cite{SARR} apply due to the constant density constraint (which is also one of the 
control parameters in the canonical ensemble). We stress that although Fisher renormalization predicts that the critical exponents are unchanged in constrained systems with logarithmic 
specific heat anomalies, as in the SARR model on the square lattice, it does predict finite-size logarithmic corrections to the scaling functions, which if neglected will lead to 
effective exponents that may differ significantly from the true asymptotic exponents of the unconstrained system. 

A very careful analysis of the criticality of the Ising magnetic lattice gas on the square lattice, in the canonical ensemble, was carried out by Ferreira and Prodanescu \cite{ALF} 
and illustrates in detail how the effective exponents depend on the scaling analysis of the constrained system. The authors point out that the values of $g_4$ at the intersection of 
the Binder cumulants for different system sizes decrease (slowly) as the system size increases and their best estimate for the cumulant at criticality is reported to be significantly larger than the corresponding 2D Ising value. Using normal scaling $\nu$ was found to differ from 2D Ising but when Fisher scaling was taken into account $\nu$ was found to approach 
the 2D Ising value \cite{ALF}. The authors also estimated $\gamma/\nu$ and obtained excellent agreement with 2D Ising when using Fisher scaling by contrast to the value obtained 
from normal scaling. The results for the 2D magnetic Ising gas show clearly that when Fisher scaling is taken into account, the effective exponents are closer to the values observed 
in the unconstrained system, as expected on theoretical grounds \cite{Fisher,ALF}. The authors stress that Fisher scaling is not a correction to normal scaling but a scaling which deviates from normal logarithmically, rendering the numerical investigation of the criticality of these systems a very challenging problem.

Finite-size-scaling theory asserts that on the critical line $T_c(\mu)$, $g_4(L)$ adopts a non-trivial value, $g_4^c$ independent of the system size $L$. For a given set of 
boundary conditions, this value of $g_4^c$ is the same for systems in the same universality class. In addition, the dependence of $g_4$ on the coupling parameter(s), $K$, in the 
critical region scales as\cite{Landau_Binder}: $(\partial g_4/\partial K) \propto L^{1/\nu}$. This is what we referred to above as normal scaling. Using normal scaling, L\'opez et al. obtained results for the constrained SARR model, consistent with $\nu=3/4$. They also report that the crossing of $g_4$ occurs at $g_4^{c} \simeq 0.638$, which is claimed to be 
the value corresponding to the $q=1$ Potts universality class (random percolation).

The use of normal scaling for the constrained SARR model, leading to L\'opez et al. conclusion of percolation critical behavior, has to be questioned. Previously \cite{SARR}, we proposed a simple argument that accounts for the effective value of $\nu = 4/3$ reported by Lopez et al. \cite{Lopez} for the constrained SARR model. We indicated that, for 
large systems, there is an additional $L/\ln L$ term in the scaling of the density derivatives compared to field derivatives. In the range of sizes investigated by \cite{Lopez} 
$L/\ln L$ is fitted by: $L /\ln L \simeq a L^{1/\nu'}$, with $\nu' \simeq 1.291$, close to $\nu = 4/3$ of the $q=1$ Potts model \cite{SARR}. This logarithmic correction arises
from the density constraint and has been discussed in much more detail by Fisher \cite{Fisher} and was investigated numerically by Ferreira and Prodanescu \cite{ALF}. We note that 
the simple $L/\ln L$ correction \cite{SARR} is in line with Fisher scaling for large systems (see equations (14) and (16) of \cite{ALF}). Therefore, in what follows we focus on the difference between the values of $g_4^c$ reported by L\'opez et al., for constrained and unconstrained SARR models. 

Lopez et al., discard the possibility of Ising criticality of the constrained SARR model based on the value of $g_4^{c}$, which differs from that of the unconstrained model: 
$g_4^{\textrm Ising} \simeq 0.611$. The results of \cite{ALF} indicate that such an assumption is far from justified. In normal scaling, appropriate for unconstrained models, finite-size scaling\cite{Stanley,Wilding} considers the singular part of the appropriate thermodynamic potential in terms of the thermodynamic fields: temperature, $T$; chemical 
potential, $\mu$; external fields. The corresponding intensive conjugate variables: energy per unit volume, density, $\rho$; magnetization are the natural variables of the 
constrained models, where Fisher scaling applies \cite{Fisher,ALF}. The SARR model on the square lattice may be described as a symmetric binary mixture, where a species corresponds 
to a given orientation. The relevant thermodynamic fields are then $T$ and  $\mu$. Within this Grand Canonical description of the SARR model (completed by taking the volume as the extensive variable that defines the system size), normal scaling theory applies. Of course, one can invesigate the criticality of the model in other ensembles but then the appropriate scaling theory must be used \cite{ALF,Desai}. 
 
In order to check the effect of the density constraint on $g_4^{c}$ in systems where $(\partial \rho/\partial \mu)_T$ diverges at the critical point we consider the behavior of the 
hard square lattice (HSL) model \cite{Guo}. The HSL  is an athermal model  (an occupied site excludes occupation of its nearest neighbor sites)
defined on the square lattice and exhibits a continuous order-disorder transition: at high densities particles occupy preferentially one of the two sublattices. The order parameter 
is defined as: $\delta = |N_1-N_2|/L^2$, where $N_i$ is the number of occupied sites in sublattice $i$, and $L^2$ is 
the number of  lattice sites. The transition of the HSL model is in the 2D Ising class and both the chemical potential and the density at the critical point are known with
high accuracy\cite{Guo}. 
 
\begin{figure}
\includegraphics[width=70mm,clip=]{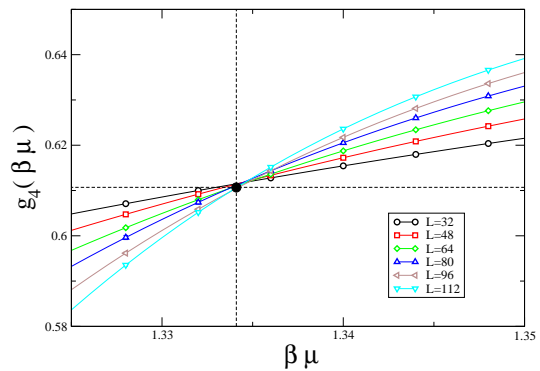}
\caption{Results for the Binder cumulant of the hard square lattice model in the grand canonical ensemble for different system sizes (See the legends).
The filled circle marks the critical chemical potential  of the HSL model and the corresponding value of $g_4^{c}$ for the 2D-Ising universality class (for unconstrained systems) with 
periodic boundary conditions.}
\label{fig1}
\end{figure}

\begin{figure}
\includegraphics[width=70mm,clip=]{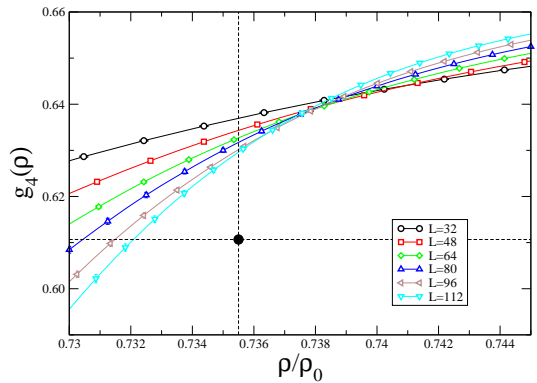}
\caption{Results for the Binder cumulant of the hard square lattice model in the canonical ensemble.
The filled circle marks the critical density of the HSL model and the corresponding value of $g_4^{c}$ for the 2D-Ising universality class (unconstrained systems) with
periodic boundary conditions. $\rho_0$ is the density at maximum lattice occupancy.}
\label{fig2}
\end{figure}

We simulated the transition of the HSL model using a multicanonical sampling procedure \cite{SARR,Lomba,Almarza2008,Almarza2009} that allows results in the canonical and grand canonical ensembles 
to be obtained simultaneously. In figures \ref{fig1} and \ref{fig2} we illustrate the results for $g_4(\beta \mu)$ and $g_4(\rho)$, in the grand canonical (unconstrained) and 
canonical (constrained) ensembles respectively. 
We find that the results in the grand canonical ensemble are fully consistent with the expected 2D Ising behavior. The curves $g_4(\beta \mu)$ for different system sizes cross 
(within error bars) at the the expected value $(\beta\mu_c,g_4^{\textrm Ising})$. However, in the canonical ensemble, the crossings occurs at a density slightly larger 
than $\rho_c$, (this could be a finite-size effect), while the crossing of $g_4$ decreases slowly as the lattice size increases, in line with the results reported for the Ising 
lattice gas model \cite{ALF}. More importantly, the results suggest that the universal value of $g_4^c$ for the constrained system may differ from the 2D Ising value for the unconstrained system, $g_4^{\textrm Ising}$. Incidentally, the crossings of $g_4$ in the canonical ensemble, occur at values close to the value reported by L\'opez et al. as the universal value of the cumulant for the $q=1$ Potts criticality. 

We conclude that the dependence of the universality class of the SARR model on the statistical ensemble, reported by L\'opez et al., is very likely the result of inadequate use 
of normal scaling to investigate the critical properties of the constrained (constant density) system. A full analysis following the lead of Ferreira and Prodanescu \cite{ALF} 
seems to be called for but it is clearly outside the scope of this Reply. 

\acknowledgments
NGA gratefully acknowledges the support from the Direcci\'on General de Investigaci\'on 
Cient\'{\i}fica  y T\'ecnica under Grant No. FIS2010-15502, and from the
Direcci\'on General de Universidades e Investigaci\'on de la Comunidad
de Madrid under Grant No. S2009/ESP-1691 and Program MODELICO-CM. MMTG and JMT acknowledge 
financial support from the Portuguese Foundation for Science and Technology (FCT) under 
Contracts nos.\ PEst-OE/FIS/UI0618/2011  and PTDC/FIS/098254/2008. MMTG acknowledges a discussion 
with Ant\'onio Lu\'\i s Ferreira.

\end{document}